\documentclass[aps,prl,floatfix,twocolumn,showpacs]{revtex4}
\usepackage{epsfig}
\usepackage{amsmath}

\begin{document}

\hsize\textwidth\columnwidth\hsize\csname@twocolumnfalse\endcsname

\title{Theory of anisotropic exchange in laterally coupled quantum dots}

\author{Fabio Baruffa$^1$, Peter Stano$^{2,3}$ and Jaroslav Fabian$^1$}
\affiliation{$^1$Institute for Theoretical Physics, University of
Regensburg, 93040 Regensburg, Germany\\
$^2$Institute of Physics, Slovak Academy of Sciences, 84511 Bratislava, Slovak Republic\\
$^3$Physics Department, University of Arizona, 1118 E 4$^{th}$ Street, Tucson, AZ 85721, USA}

\vskip1.5truecm
\begin{abstract}
The effects of spin-orbit coupling on the two-electron spectra in lateral coupled quantum dots are
investigated analytically and numerically. It is demonstrated that in the absence of magnetic field the 
exchange interaction is practically unaffected by spin-orbit coupling, for any interdot coupling,
boosting prospects for spin-based quantum computing. The anisotropic exchange appears at finite
magnetic fields. A numerically accurate effective spin Hamiltonian
for modeling spin-orbit-induced two-electron spin dynamics in the presence of magnetic field
is proposed.
\end{abstract}
\pacs{71.70.Gm, 71.70.Ej, 73.21.La, 75.30.Et} 

\maketitle

The electron spins in quantum dots are natural and viable qubits for quantum computing,\cite{loss1998:PRA}
as evidenced by the impressive recent experimental progress \cite{hanson2007:RMP,taylor2007:PRB}
in spin detection and spin relaxation,\cite{elzerman2004:N,koppens2008:PRL} as well as in coherent spin manipulation.\cite{petta2005:S,nowack2007:S}
In coupled dots, the two-qubit quantum gates are realized by manipulating the exchange coupling which originates in the Coulomb interaction and the Pauli principle.\cite{loss1998:PRA,hu2000:PRA} How is the exchange modified by the presence of the spin-orbit coupling?
In general, the usual (isotropic) exchange changes its magnitude while a new, functionally different
form of exchange, called anisotropic, appears, breaking the spin-rotational symmetry.
Such changes are a nuisance from the perspective of the error correction,\cite{stepanenko2003:PRB} although the anisotropic exchange could also induce quantum gating.\cite{stepanenko2004:PRL,zhao2006:PRB}

The anisotropic exchange of coupled localized electrons has a convoluted history\cite{gangadharaiah2008:PRL,shekhtman1992:PRL,zheludev1999:PRB,tserkovnyak2009:PRL,chutia2006:PRB,gorkov2003:PRBb,kunikeev2008:PRB}. The question boils down to determining
the leading order in which the spin-orbit coupling affects both the isotropic and anisotropic
exchange. At zero magnetic field, the second order was suggested,\cite{kavokin2001:PRB} with later revisions showing the effects are absent in the second order.\cite{kavokin2004:PRB,gangadharaiah2008:PRL} The analytical complexities make a numerical analysis particularly useful.

Here we perform numerically exact calculations of the isotropic and anisotropic
exchange in realistic GaAs coupled quantum dots in the presence of both the Dresselhaus and
Bychkov-Rashba spin-orbit interactions.\cite{fabian2007:APS} The numerics allows
us to make authoritative statements about the exchange. We establish that in zero magnetic field the second-order spin-orbit effects are absent at {\it all} interdot couplings. Neither is the isotropic exchange affected, nor is the anisotropic
exchange present. At finite magnetic fields the anisotropic coupling appears. We derive
a spin-exchange Hamiltonian describing this behavior, generalizing the existing descriptions; we do not rely on weak coupling approximations such as the Heitler-London one.
The model is proven
highly accurate by comparison with our numerics and we propose it as a realistic  effective
model for the two-spin dynamics in coupled quantum dots.


Our microscopic description is the single band effective mass envelope function approximation; we neglect multiband effects.\cite{badescu2005:PRB,glazov2009:PRB} We consider a two electron double dot whose lateral confinement is defined electrostatically by metallic gates on the top of a semiconductor heterostructure. The heterostructure, grown along [001] direction, provides strong perpendicular confinement, such that electrons are strictly two dimensional, with the Hamiltonian (subscript $i$ labels the electrons)
\begin{equation}
H=\sum_{i=1,2} \left( T_i + V_i + H_{Z,i} + H_{{\rm so},i} \right) + H_C.
\label{eq:two electron H}
\end{equation}
The single electron terms are the kinetic energy, model confinement potential, and the Zeeman term,
\begin{eqnarray}
T&=&{\bf P}^2/2m = (-{\rm i}\hbar \boldsymbol{\nabla}+e{\bf A})^2/ 2m,\\
V&=&(1/2) m \omega^2 [{\rm min}\{(x-d)^2,(x+d)^2\} + y^2],\\
H_Z&=&(g/2)(e\hbar/2m_e) {\bf B}\cdot \boldsymbol{\sigma} =  \mu {\bf B}\cdot \boldsymbol{\sigma},
\end{eqnarray}
and spin-orbit interactions---linear and cubic Dresselhaus, and Bychkov-Rashba\cite{fabian2007:APS},
\begin{eqnarray}
H_d&=&(\hbar/m l_d) (-\sigma_x P_x + \sigma_y P_y),\\
H_{d3}&=&(\gamma_c/2\hbar^3) (\sigma_x P_x P_y^2 - \sigma_y P_y P_x^2)+\textrm{Herm. conj.},\\
H_{br}&=&(\hbar/m l_{br}) (\sigma_x P_y - \sigma_y P_x),
\end{eqnarray}
which we lump together as $H_{\rm so}={\bf w} \cdot \boldsymbol{\sigma}$. The position ${\bf r}$ and momentum ${\bf P}$ vectors are two dimensional (in-plane); $m/m_e$ is the effective/electron mass, $e$ is the proton charge, ${\bf A}=B_z(-y,x)/2$ is the in-plane  vector potential to magnetic field ${\bf B}=(B_x,B_y,B_z)$, $g$ is the electron g-factor, $\boldsymbol{\sigma}$ are Pauli matrices, and $\mu$ is the renormalized magnetic moment. The double dot confinement is modeled by two equal single dots displaced along [100] by $\pm d$, each with a harmonic potential with confinement energy $\hbar \omega$. The spin-orbit interactions are parametrized by the bulk material constant $\gamma_c$ and the heterostructure dependent spin-orbit lengths $l_{br}$, $l_d$. Finally, the Coulomb interaction is
$H_C=(e^2/4\pi \epsilon) |{\bf r_1}-{\bf r_2}|^{-1}$, with the dielectric constant $\epsilon$. 

The numerical results are obtained by exact diagonalization (configuration interaction method). The two electron Hamiltonian is diagonalized in the basis of Slater determinants constructed from numerical single electron states in the double dot potential. Typically we use 21 single electron states, resulting in the relative error for energies of order $10^{-5}$. We use material parameters of GaAs: $m=0.067 m_e$, $g=-0.44$, $\gamma_c=27.5$ meV\AA$^3$, a typical single dot confinement energy $\hbar \omega=1.1$ meV, and spin-orbit lengths $l_d=1.26\,\mu$m and $l_{br}=1.72\,\mu$m from a fit to a spin relaxation experiment.\cite{stano2006:PRL,stano2006:PRB}

\begin{figure}
\centerline{\psfig{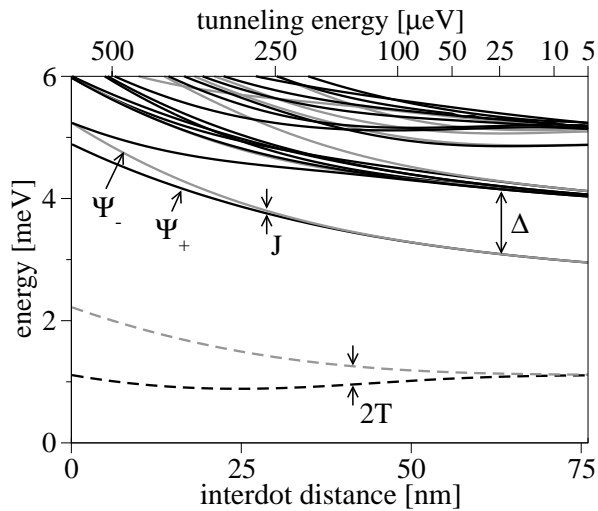}}
\caption{Calculated double dot spectrum as a function of the interdot distance/tunneling energy. Spin is not considered and the magnetic field is zero. Solid lines show the two electron energies. The two lowest states are explicitly labeled, split by the isotropic exchange $J$ and displaced from the nearest higher excited state by $\Delta$. For comparison, the two lowest single electron states are shown (dashed), split by twice the tunneling energy $T$. State spatial symmetry is denoted by darker (symmetric) and lighter (antisymmetric) lines. }
\label{fig:spectrum}
\end{figure}

Let us first neglect the spin and look at the spectrum in zero magnetic field as a function of the interdot distance ($2d$)/tunneling energy, Fig.~\ref{fig:spectrum}. At $d=0$ our model describes a single dot. The interdot coupling gets weaker as one moves to the right; both the isotropic exchange $J$ and the tunneling energy $T$ decay exponentially. The symmetry of the confinement potential assures the electron wavefunctions are symmetric or antisymmetric upon inversion. The  two lowest states, $\Psi_\pm$, are separated from the higher excited states by an appreciable gap $\Delta$, what justifies the restriction to the two lowest orbital wavefunctions for the spin qubit pair at a weak coupling. Our further derivations are based on the observation
\begin{equation}
P \Psi_\pm = \pm \Psi_\pm,\qquad I_1 I_2 \Psi_\pm = \pm \Psi_\pm,
\label{eq:symmetry}
\end{equation}
where $I f(x,y)=f(-x,-y)$ is the inversion operator and $P f_1 g_2 = f_2 g_1$ is the particle exchange operator. Functions $\Psi_\pm$ in the Heitler-London approximation fulfill Eq.~(\ref{eq:symmetry}). However, unlike Heitler-London, Eq.~(\ref{eq:symmetry}) is valid generally in symmetric double dots, as we learn from numerics (we saw it valid in all cases we studied).

Let us reinstate the spin. The restricted two qubit subspace amounts to the following four states ($S$ stands for singlet, $T$ for triplet),
\begin{equation}
\{ \Phi_i \}_{i=1,\ldots,4}=\{ \Psi_+ S, \Psi_- T_+, \Psi_- T_0, \Psi_- T_-  \},
\label{eq:basis}
\end{equation}
Within this basis, the system is described by a 4 by 4 Hamiltonian with matrix elements $(H_4)_{ij}=\langle \Phi_i |H| \Phi_j\rangle$. Without spin-orbit interactions, this Hamiltonian is diagonal, with the singlet and triplets split by the isotropic exchange $J$,\cite{loss1998:PRA,hu2000:PRA} and the triplets split by the Zeeman energy $\mu B$. It is customary to refer only to the spinor part of the basis states, using the sigma matrices, resulting in the isotropic exchange Hamiltonian,
\begin{equation}
H_{\rm iso} = (J/4) \boldsymbol{\sigma_1} \cdot \boldsymbol{\sigma_2} + \mu {\bf B} \cdot  (\boldsymbol{\sigma_1} + \boldsymbol{\sigma_2}).
\label{eq:isotropic exchange}
\end{equation}

A naive approach to include the spin-orbit interaction is to consider it within the basis of Eq.~(\ref{eq:basis}). This gives the Hamiltonian $H_{\rm ex}^\prime=H_{\rm iso}+H_{\rm aniso}^\prime$, where
\begin{equation}
H_{\rm aniso}^\prime={\bf a^\prime}\cdot (\boldsymbol{\sigma_1} - \boldsymbol{\sigma_2}) + {\bf b^\prime}\cdot (\boldsymbol{\sigma_1} \times \boldsymbol{\sigma_2}),
\label{eq:anisotropic exchange}
\end{equation}
with the six real parameters given by spin-orbit vectors
\begin{eqnarray}
{\bf a^\prime} = {\rm Re} \langle \Psi_+ | {\bf w_1} | \Psi_- \rangle, \quad {\bf b^\prime} = {\rm Im} \langle \Psi_+ | {\bf w_1} | \Psi_- \rangle.
\label{eq:spin orbit vectors}
\end{eqnarray}
The form of the Hamiltonian follows solely from the inversion symmetry $I {\bf w} = -{\bf w}$ and Eq.~(\ref{eq:symmetry}). The spin-orbit coupling appears in the first order.

\begin{figure}
\centerline{\psfig{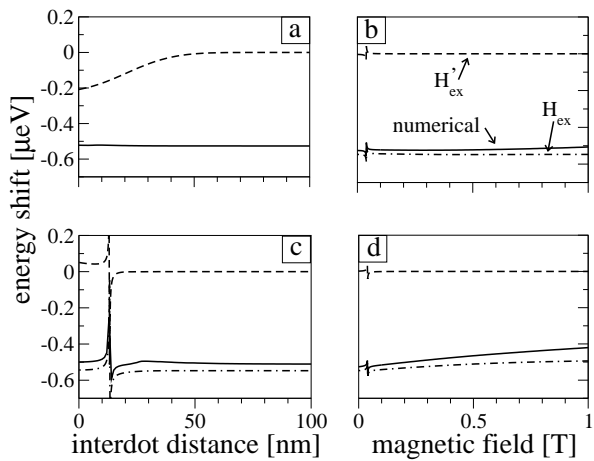}}
\caption{The spin-orbit induced energy shift as a function of the interdot distance (left) and perpendicular magnetic field (right). a) Singlet in zero magnetic field, c) singlet at 1 Tesla field, b) and d) singlet and triplet $T_+$ at the interdot distance 55 nm corresponding to the zero field isotropic exchange of 1 $\mu$eV. The exchange models $H_{\rm ex}^\prime$ (dashed) and $H_{\rm ex}$ (dot-dashed) are compared with the numerics (solid).
}
\label{fig:socont}
\end{figure}

The Hamiltonian $H_{\rm ex}^\prime$ fares badly with numerics. Figure \ref{fig:socont} shows the energy shifts caused by the spin-orbit coupling for selected states, at different interdot couplings and perpendicular magnetic fields.  The model is completely off even though we use numerical wavefunctions $\Psi_\pm$ in Eq.~(\ref{eq:spin orbit vectors}) without further approximations.

To improve the analytical model, we remove the linear spin-orbit terms from the Hamiltonian using transformation\cite{aleiner2001:PRL,levitov2003:PRB,kavokin2004:PRB}
\begin{equation}
 U=\exp\left[-({\rm i}/2){\bf n_1} \cdot \boldsymbol{\sigma_1}
-({\rm i}/2){\bf n_2} \cdot \boldsymbol{\sigma_2}\right],
\end{equation}
where ${\bf n} = \left( x/l_d-y/l_{br}, x/l_{br}-y/l_d,0\right)$.

Up to the second order in small quantities (the spin-orbit and Zeeman interactions), the transformed Hamiltonian $\overline{H}= U H U^\dagger$ is the same as the original, Eq.~(\ref{eq:two electron H}), except for the linear spin-orbit interactions:
\begin{equation}
\overline{H}_{\rm so} = -(\mu {\bf B}\times {\bf n}) \cdot \boldsymbol{\sigma} + (K_-/\hbar) L_z \sigma_z - K_+ ,
\label{eq:effective so}
\end{equation}
where $K_\pm = (\hbar^2/4 m l_d^2) \pm (\hbar^2/4 m l_{br}^2)$. In the unitarily transformed basis, we again restrict the Hilbert space to the lowest four states, getting 
the effective Hamiltonian
\begin{equation} 
\begin{split}
H_{\rm ex}&=(J/4) \boldsymbol{\sigma_1} \cdot \boldsymbol{\sigma_2} + \mu ({\bf B}+{\bf B}_{\rm so})  \cdot (\boldsymbol{\sigma_1} + \boldsymbol{\sigma_2})\\
& +{\bf a} \cdot (\boldsymbol{\sigma_1} - \boldsymbol{\sigma_2}) + {\bf b} \cdot (\boldsymbol{\sigma_1} \times \boldsymbol{\sigma_2})-2K_+.
\end{split}
\label{eq:final result}
\end{equation}
The operational form is the same as for $H_{\rm ex}^\prime$. The qualitative difference is in the
way the spin-orbit enters the parameters. First, a contribution to the
Zeeman term,
\begin{equation}
\mu {\bf B}_{\rm so} = {\bf \hat{z}} (K_-/\hbar)  \langle \Psi_- | L_{z,1} | \Psi_- \rangle,
\end{equation}
appears due to the inversion symmetric part of Eq.~(\ref{eq:effective so}). Second, the
spin-orbit vectors are linearly proportional to both the spin-orbit coupling and
magnetic field,
\begin{subequations}
\label{eq:spin orbit vectors 2}
\begin{eqnarray}
{\bf a} &=& -\mu {\bf B} \times {\rm Re} \langle \Psi_+ | {\bf n_1} | \Psi_- \rangle,\\
{\bf b} &=& -\mu {\bf B} \times {\rm Im} \langle \Psi_+ | {\bf n_1} | \Psi_- \rangle.
\end{eqnarray}
\end{subequations}
The effective model and the exact data agree very well for all interdot couplings, as seen in Fig.~\ref{fig:socont}.

At zero magnetic field, only the first and the last term in Eq.~(\ref{eq:final result}) survive. This is the result of Ref.~\cite{kavokin2004:PRB}, where primed operators were used to refer to the fact that the Hamiltonian $H_{ex}$ refers to the transformed basis, $\{U\Phi_i\}$. Note that if a basis separable in orbital and spin part is required, undoing $U$ necessarily yields the original Hamiltonian Eq.~(\ref{eq:two electron H}), and the restriction to the four lowest states gives $H_{ex}^\prime$. Replacing the coordinates $(x,y)$ by mean values $(\pm d, 0)$\cite{gangadharaiah2008:PRL} visualizes the Hamiltonian $H_{ex}$ as an interaction through rotated sigma matrices, but this is just an approximation, valid if $d, l_{so} \gg l_0$.

\begin{figure}
\centerline{\psfig{file=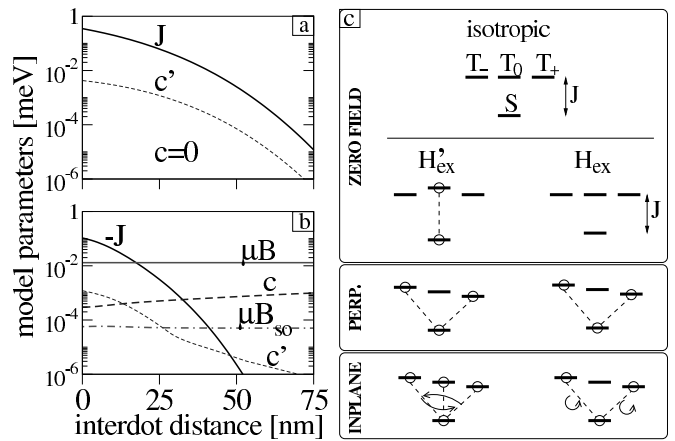,width=\linewidth}}
\caption{ a) The isotropic and anisotropic exchange as functions of the interdot distance at zero magnetic field. b) The isotropic exchange $J$, anisotropic exchange $c$/$c^\prime$, the Zeeman splitting $\mu B$, and its spin-orbit part
$\mu B_{so}$ at perpendicular magnetic field of 1 Tesla. c) Schematics of the exchange-split
four lowest states for the three models, $H_{\rm iso}$, $H_{\rm ex}^\prime$, and $H_{\rm ex}$, which include the spin-orbit coupling in no, first, and second order, respectively, at zero
magnetic field (top). The latter two models are compared in perpendicular
and in-plane magnetic fields as well. The eigenenergies are indicated by the solid lines. The dashed lines show which states are coupled by the spin-orbit coupling. The arrows indicate the redistribution
of the couplings as the in-plane field direction changes with respect to the crystallographic axes (see the
main text).}
\label{fig:params}
\end{figure}

One of our main numerical results is establishing the validity of the Hamiltonian in Eq.~(\ref{eq:final result}) for $B=0$, confirming recent analytic predictions and extending their applicability beyond the weak coupling limit.
{\it In the transformed basis, the spin-orbit interactions do not lead to any anisotropic exchange, nor do they modify the isotropic one.} In fact, this result could have been
anticipated from its single-electron analog: at zero magnetic field there is no spin-orbit contribution to the tunneling energy,\cite{stano2005:PRB} going opposite to the intuitive notion of the spin-orbit coupling induced coherent spin rotation and spin-flip tunneling amplitudes. 
Figure \ref{fig:params}a summarizes this case, with the isotropic exchange as the only nonzero parameter of model $H_{\rm ex}$. In contrast, model $H_{\rm ex}^\prime$ predicts a finite anisotropic exchange.\footnote{The spin-orbit vectors are determined up to the relative phase of states $\Psi_+$ and $\Psi_-$. The observable quantity is $c=\surd a^2 + b^2$ and analogously for $c^\prime=\surd a^{\prime 2} + b^{\prime 2}$.}

From the concept of dressed qubits\cite{wu2003:PRL} it follows that the main consequence of the spin-orbit interaction, the transformation $U$ of the basis, is not a nuisance for quantum computation. We expect this property to hold also for a qubit array, since the electrons are at fixed positions without the possibility of a long distance tunneling. However, a rigorous analysis of this point is beyond the scope of this article. If electrons are allowed to move, $U$ results in the spin relaxation.\cite{kavokin2008:SST}

Figure \ref{fig:params}b shows model parameters in 1 Tesla perpendicular magnetic field. 
The isotropic exchange again decays exponentially.  As it becomes smaller than the Zeeman energy, the singlet state anticrosses one of the polarized triplets (seen as cusps on Fig.~\ref{fig:socont}). Here it is $T_+$, due to the negative sign of both the isotropic exchange and the g-factor. 
Because the Zeeman energy always dominates the spin-dependent terms and the singlet and triplet $T_0$ are never coupled (see below), the anisotropic exchange influences the energy in the second order.\cite{gangadharaiah2008:PRL}
Note the difference in the strengths. In 
$H_{\rm ex}^\prime$ the anisotropic exchange falls off exponentially, while $H_{\rm ex}$ predicts non-exponential behavior, resulting in spin-orbit effects larger by orders of magnitude. The effective magnetic field $B_{\rm so}$ is always much smaller than the real magnetic field and can be neglected in most cases.

Figure \ref{fig:params}c compares analytical models. In zero field and no spin-orbit interactions, the isotropic exchange Hamiltonian $H_{\rm iso}$ describes the system. Including the spin-orbit coupling in the first order, $H_{\rm ex}^\prime$, gives a nonzero coupling between the singlet and triplet $T_0$. Going to the second order, the effective model $H_{\rm ex}$ shows there are no spin-orbit effects (other than the basis redefinition).

The Zeeman interaction splits the three triplets in a finite magnetic field. Both $H_{\rm ex}^\prime$ and $H_{\rm ex}$ predict the same type of coupling in a perpendicular field, between the singlet and the two polarized triplets. Interestingly, in in-plane fields the two models differ qualitatively. In $H_{\rm ex}^\prime$ the spin-orbit vectors are fixed in the plane. Rotation of the magnetic field ``redistributes'' the couplings among the triplets. (This anisotropy with respect to the crystallographic  axis is due to the $C_{2v}$ symmetry of the two-dimensional electron gas in GaAs, imprinted in the Bychkov-Rashba and Dresselhaus interactions.\cite{fabian2007:APS}) In contrast, the spin-orbit vectors of $H_{\rm ex}$ are always perpendicular to the magnetic field. Remarkably, aligning the magnetic field along a special direction (here we allow an arbitrary positioned dot, with $\delta$ the angle between the main dot axis and the crystallographic $x$ axis), 
\begin{equation}
[l_{br}-l_d \tan \delta, l_d-l_{br}\tan\delta,0],
\label{eq:magic direction}
\end{equation}
{\it all the spin-orbit effects disappear once again}, as if $B$ were zero. (An analogous angle was reported for a single dot in Ref.~\cite{golovach2008:PRB}). This has strong implications for the spin-orbit induced singlet-triplet relaxation. Indeed, $S\leftrightarrow T_0$ transitions are {\it ineffective at any magnetic field}, as these two states are never coupled in our model. Second, $S \leftrightarrow T_\pm$ transitions will show strong (orders of magnitude) anisotropy with respect to the field direction, reaching minimum at the direction given by Eq.~(\ref{eq:magic direction}). This prediction is straightforwardly testable in experiments on two electron spin relaxation.

Our derivation was based on the inversion symmetry of the potential only. What are the limits of our model? We neglected third order terms in $\overline{H}_{so}$ and, restricting the Hilbert space, corrections from higher excited orbital states. (Among the latter is the non-exponential spin-spin coupling\cite{gangadharaiah2008:PRL}). Compared to the second order terms we keep, these are smaller by (at least) $d/l_{so}$ and $c/\Delta$, respectively.[35]  Apart from the analytical estimates, the numerics, which includes all terms, assures us that both of these are negligible. Based on numerics we also conclude our analytical model stays quantitatively faithful even at the strong coupling limit, where $\Delta \to 0$.
More involved is the influence of the cubic Dresselhaus term, which is not removed by the unitary transformation. This term is the main source for the discrepancy of the model and the numerical data in finite fields. Most importantly, it does not change our results for $B=0$.

Concluding, we studied the effects of spin-orbit coupling on the exchange in lateral coupled GaAs quantum dots. We 
 derive and support by precise numerics an effective Hamiltonian for two spin qubits, generalizing the existing models. The effective anisotropic exchange model should be useful in precise analysis of the
 physical realizations of quantum computing schemes based on quantum dot spin qubits, as well as in the physics
 of electron spins in quantum dots in general. Our analysis should also improve the current understanding 
 of the singlet-triplet spin relaxation \cite{shen2007:PRB,sherman2005:PRB,climente2007:PRB,olendski2007:PRB}.

This work was supported by DFG GRK 638, SPP 1285, NSF grant DMR-0706319, RPEU-0014-06, ERDF OP R\&D ``QUTE'', CE SAS QUTE and DAAD.

\bibliography{../../references/quantum_dot}

\begin{thebibliography}{35}
\expandafter\ifx\csname natexlab\endcsname\relax\def\natexlab#1{#1}\fi
\expandafter\ifx\csname bibnamefont\endcsname\relax
  \def\bibnamefont#1{#1}\fi
\expandafter\ifx\csname bibfnamefont\endcsname\relax
  \def\bibfnamefont#1{#1}\fi
\expandafter\ifx\csname citenamefont\endcsname\relax
  \def\citenamefont#1{#1}\fi
\expandafter\ifx\csname url\endcsname\relax
  \def\url#1{\texttt{#1}}\fi
\expandafter\ifx\csname urlprefix\endcsname\relax\def\urlprefix{URL }\fi
\providecommand{\bibinfo}[2]{#2}
\providecommand{\eprint}[2][]{\url{#2}}

\bibitem[{\citenamefont{Loss and DiVincenzo}(1998)}]{loss1998:PRA}
\bibinfo{author}{\bibfnamefont{D.}~\bibnamefont{Loss}} \bibnamefont{and}
  \bibinfo{author}{\bibfnamefont{D.~P.} \bibnamefont{DiVincenzo}},
  \bibinfo{journal}{Phys. Rev. A} \textbf{\bibinfo{volume}{57}},
  \bibinfo{pages}{120} (\bibinfo{year}{1998}).

\bibitem[{\citenamefont{Hanson et~al.}(2007)\citenamefont{Hanson, Kouwenhoven,
  Petta, Tarucha, and Vandersypen}}]{hanson2007:RMP}
\bibinfo{author}{\bibfnamefont{R.}~\bibnamefont{Hanson}},
  \bibinfo{author}{\bibfnamefont{L.~P.} \bibnamefont{Kouwenhoven}},
  \bibinfo{author}{\bibfnamefont{J.~R.} \bibnamefont{Petta}},
  \bibinfo{author}{\bibfnamefont{S.}~\bibnamefont{Tarucha}}, \bibnamefont{and}
  \bibinfo{author}{\bibfnamefont{L.~M.~K.} \bibnamefont{Vandersypen}},
  \bibinfo{journal}{Rev. Mod. Phys.} \textbf{\bibinfo{volume}{79}},
  \bibinfo{pages}{1217} (\bibinfo{year}{2007}).

\bibitem[{\citenamefont{Taylor et~al.}(2007)\citenamefont{Taylor, Petta,
  Johnson, Yacoby, Marcus, and Lukin}}]{taylor2007:PRB}
\bibinfo{author}{\bibfnamefont{J.~M.} \bibnamefont{Taylor}},
  \bibinfo{author}{\bibfnamefont{J.~R.} \bibnamefont{Petta}},
  \bibinfo{author}{\bibfnamefont{A.~C.} \bibnamefont{Johnson}},
  \bibinfo{author}{\bibfnamefont{A.}~\bibnamefont{Yacoby}},
  \bibinfo{author}{\bibfnamefont{C.~M.} \bibnamefont{Marcus}},
  \bibnamefont{and} \bibinfo{author}{\bibfnamefont{M.~D.} \bibnamefont{Lukin}},
  \bibinfo{journal}{Phys. Rev. B} \textbf{\bibinfo{volume}{76}},
  \bibinfo{pages}{035315} (\bibinfo{year}{2007}).

\bibitem[{\citenamefont{Elzerman et~al.}(2004)\citenamefont{Elzerman, Hanson,
  {Willems van Beveren}, Witkamp, Vandersypen, and
  Kouwenhoven}}]{elzerman2004:N}
\bibinfo{author}{\bibfnamefont{J.~M.} \bibnamefont{Elzerman}},
  \bibinfo{author}{\bibfnamefont{R.}~\bibnamefont{Hanson}},
  \bibinfo{author}{\bibfnamefont{L.~H.} \bibnamefont{{Willems van Beveren}}},
  \bibinfo{author}{\bibfnamefont{B.}~\bibnamefont{Witkamp}},
  \bibinfo{author}{\bibfnamefont{L.~M.~K.} \bibnamefont{Vandersypen}},
  \bibnamefont{and} \bibinfo{author}{\bibfnamefont{L.~P.}
  \bibnamefont{Kouwenhoven}}, \bibinfo{journal}{Nature}
  \textbf{\bibinfo{volume}{430}}, \bibinfo{pages}{431} (\bibinfo{year}{2004}).

\bibitem[{\citenamefont{Koppens et~al.}(2008)\citenamefont{Koppens, Nowack, and
  Vandersypen}}]{koppens2008:PRL}
\bibinfo{author}{\bibfnamefont{F.~H.~L.} \bibnamefont{Koppens}},
  \bibinfo{author}{\bibfnamefont{K.~C.} \bibnamefont{Nowack}},
  \bibnamefont{and} \bibinfo{author}{\bibfnamefont{L.~M.~K.}
  \bibnamefont{Vandersypen}}, \bibinfo{journal}{Phys. Rev. Lett.}
  \textbf{\bibinfo{volume}{100}}, \bibinfo{pages}{236802}
  (\bibinfo{year}{2008}).

\bibitem[{\citenamefont{Petta et~al.}(2005)\citenamefont{Petta, Johnson,
  Taylor, Laird, Yacoby, Lukin, Marcus, Hanson, and Gossard}}]{petta2005:S}
\bibinfo{author}{\bibfnamefont{J.~R.} \bibnamefont{Petta}},
  \bibinfo{author}{\bibfnamefont{A.~C.} \bibnamefont{Johnson}},
  \bibinfo{author}{\bibfnamefont{J.~M.} \bibnamefont{Taylor}},
  \bibinfo{author}{\bibfnamefont{E.~A.} \bibnamefont{Laird}},
  \bibinfo{author}{\bibfnamefont{A.}~\bibnamefont{Yacoby}},
  \bibinfo{author}{\bibfnamefont{M.~D.} \bibnamefont{Lukin}},
  \bibinfo{author}{\bibfnamefont{C.~M.} \bibnamefont{Marcus}},
  \bibinfo{author}{\bibfnamefont{M.~P.} \bibnamefont{Hanson}},
  \bibnamefont{and} \bibinfo{author}{\bibfnamefont{A.~C.}
  \bibnamefont{Gossard}}, \bibinfo{journal}{Science}
  \textbf{\bibinfo{volume}{309}}, \bibinfo{pages}{2180} (\bibinfo{year}{2005}).

\bibitem[{\citenamefont{Nowack et~al.}(2007)\citenamefont{Nowack, Koppens,
  Nazarov, and Vandersypen}}]{nowack2007:S}
\bibinfo{author}{\bibfnamefont{K.~C.} \bibnamefont{Nowack}},
  \bibinfo{author}{\bibfnamefont{F.~H.~L.} \bibnamefont{Koppens}},
  \bibinfo{author}{\bibfnamefont{Y.~V.} \bibnamefont{Nazarov}},
  \bibnamefont{and} \bibinfo{author}{\bibfnamefont{L.~M.~K.}
  \bibnamefont{Vandersypen}}, \bibinfo{journal}{Science}
  \textbf{\bibinfo{volume}{318}}, \bibinfo{pages}{1430} (\bibinfo{year}{2007}).

\bibitem[{\citenamefont{Hu and {Das Sarma}}(2000)}]{hu2000:PRA}
\bibinfo{author}{\bibfnamefont{X.}~\bibnamefont{Hu}} \bibnamefont{and}
  \bibinfo{author}{\bibfnamefont{S.}~\bibnamefont{{Das Sarma}}},
  \bibinfo{journal}{Phys. Rev. A} \textbf{\bibinfo{volume}{61}},
  \bibinfo{pages}{062301} (\bibinfo{year}{2000}).

\bibitem[{\citenamefont{Stepanenko et~al.}(2003)\citenamefont{Stepanenko,
  Bonesteel, DiVincenzo, Burkard, and Loss}}]{stepanenko2003:PRB}
\bibinfo{author}{\bibfnamefont{D.}~\bibnamefont{Stepanenko}},
  \bibinfo{author}{\bibfnamefont{N.~E.} \bibnamefont{Bonesteel}},
  \bibinfo{author}{\bibfnamefont{D.~P.} \bibnamefont{DiVincenzo}},
  \bibinfo{author}{\bibfnamefont{G.}~\bibnamefont{Burkard}}, \bibnamefont{and}
  \bibinfo{author}{\bibfnamefont{D.}~\bibnamefont{Loss}},
  \bibinfo{journal}{Phys. Rev. B} \textbf{\bibinfo{volume}{68}},
  \bibinfo{pages}{115306} (\bibinfo{year}{2003}).

\bibitem[{\citenamefont{Stepanenko and Bonesteel}(2004)}]{stepanenko2004:PRL}
\bibinfo{author}{\bibfnamefont{D.}~\bibnamefont{Stepanenko}} \bibnamefont{and}
  \bibinfo{author}{\bibfnamefont{N.~E.} \bibnamefont{Bonesteel}},
  \bibinfo{journal}{Phys. Rev. Lett.} \textbf{\bibinfo{volume}{93}},
  \bibinfo{pages}{140501} (\bibinfo{year}{2004}).

\bibitem[{\citenamefont{Zhao et~al.}(2006)\citenamefont{Zhao, Zhong, Zhu, and
  Sun}}]{zhao2006:PRB}
\bibinfo{author}{\bibfnamefont{N.}~\bibnamefont{Zhao}},
  \bibinfo{author}{\bibfnamefont{L.}~\bibnamefont{Zhong}},
  \bibinfo{author}{\bibfnamefont{J.-L.} \bibnamefont{Zhu}}, \bibnamefont{and}
  \bibinfo{author}{\bibfnamefont{C.~P.} \bibnamefont{Sun}},
  \bibinfo{journal}{Phys. Rev. B} \textbf{\bibinfo{volume}{74}},
  \bibinfo{pages}{075307} (\bibinfo{year}{2006}).

\bibitem[{\citenamefont{Gangadharaiah et~al.}(2008)\citenamefont{Gangadharaiah,
  Sun, and Starykh}}]{gangadharaiah2008:PRL}
\bibinfo{author}{\bibfnamefont{S.}~\bibnamefont{Gangadharaiah}},
  \bibinfo{author}{\bibfnamefont{J.}~\bibnamefont{Sun}}, \bibnamefont{and}
  \bibinfo{author}{\bibfnamefont{O.~A.} \bibnamefont{Starykh}},
  \bibinfo{journal}{Phys. Rev. Lett.} \textbf{\bibinfo{volume}{100}},
  \bibinfo{pages}{156402} (\bibinfo{year}{2008}).

\bibitem[{\citenamefont{Shekhtman et~al.}(1992)\citenamefont{Shekhtman,
  Entin-Wohlman, and Aharony}}]{shekhtman1992:PRL}
\bibinfo{author}{\bibfnamefont{L.}~\bibnamefont{Shekhtman}},
  \bibinfo{author}{\bibfnamefont{O.}~\bibnamefont{Entin-Wohlman}},
  \bibnamefont{and} \bibinfo{author}{\bibfnamefont{A.}~\bibnamefont{Aharony}},
  \bibinfo{journal}{Phys. Rev. Lett.} \textbf{\bibinfo{volume}{69}},
  \bibinfo{pages}{836} (\bibinfo{year}{1992}).

\bibitem[{\citenamefont{Zheludev et~al.}(1999)\citenamefont{Zheludev, Maslov,
  Shirane, Tsukada, Masuda, Uchinokura, Zaliznyak, Erwin, and
  Regnault}}]{zheludev1999:PRB}
\bibinfo{author}{\bibfnamefont{A.}~\bibnamefont{Zheludev}},
  \bibinfo{author}{\bibfnamefont{S.}~\bibnamefont{Maslov}},
  \bibinfo{author}{\bibfnamefont{G.}~\bibnamefont{Shirane}},
  \bibinfo{author}{\bibfnamefont{I.}~\bibnamefont{Tsukada}},
  \bibinfo{author}{\bibfnamefont{T.}~\bibnamefont{Masuda}},
  \bibinfo{author}{\bibfnamefont{K.}~\bibnamefont{Uchinokura}},
  \bibinfo{author}{\bibfnamefont{I.}~\bibnamefont{Zaliznyak}},
  \bibinfo{author}{\bibfnamefont{R.}~\bibnamefont{Erwin}}, \bibnamefont{and}
  \bibinfo{author}{\bibfnamefont{L.~P.} \bibnamefont{Regnault}},
  \bibinfo{journal}{Phys. Rev. B} \textbf{\bibinfo{volume}{59}},
  \bibinfo{pages}{11432} (\bibinfo{year}{1999}).

\bibitem[{\citenamefont{Tserkovnyak and
  Kindermann}(2009)}]{tserkovnyak2009:PRL}
\bibinfo{author}{\bibfnamefont{Y.}~\bibnamefont{Tserkovnyak}} \bibnamefont{and}
  \bibinfo{author}{\bibfnamefont{M.}~\bibnamefont{Kindermann}},
  \bibinfo{journal}{Phys. Rev. Lett.} \textbf{\bibinfo{volume}{102}},
  \bibinfo{pages}{126801} (\bibinfo{year}{2009}).

\bibitem[{\citenamefont{Chutia et~al.}(2006)\citenamefont{Chutia, Friesen, and
  Joynt}}]{chutia2006:PRB}
\bibinfo{author}{\bibfnamefont{S.}~\bibnamefont{Chutia}},
  \bibinfo{author}{\bibfnamefont{M.}~\bibnamefont{Friesen}}, \bibnamefont{and}
  \bibinfo{author}{\bibfnamefont{R.}~\bibnamefont{Joynt}},
  \bibinfo{journal}{Phys. Rev. B} \textbf{\bibinfo{volume}{73}},
  \bibinfo{pages}{241304(R)} (\bibinfo{year}{2006}).

\bibitem[{\citenamefont{Gorkov and Krotkov}(2003)}]{gorkov2003:PRBb}
\bibinfo{author}{\bibfnamefont{L.~P.} \bibnamefont{Gorkov}} \bibnamefont{and}
  \bibinfo{author}{\bibfnamefont{P.~L.} \bibnamefont{Krotkov}},
  \bibinfo{journal}{Phys. Rev. B} \textbf{\bibinfo{volume}{67}},
  \bibinfo{pages}{033203} (\bibinfo{year}{2003}).

\bibitem[{\citenamefont{Kunikeev and Lidar}(2008)}]{kunikeev2008:PRB}
\bibinfo{author}{\bibfnamefont{S.~D.} \bibnamefont{Kunikeev}} \bibnamefont{and}
  \bibinfo{author}{\bibfnamefont{D.~A.} \bibnamefont{Lidar}},
  \bibinfo{journal}{Phys. Rev. B} \textbf{\bibinfo{volume}{77}},
  \bibinfo{pages}{045320} (\bibinfo{year}{2008}).

\bibitem[{\citenamefont{Kavokin}(2001)}]{kavokin2001:PRB}
\bibinfo{author}{\bibfnamefont{K.~V.} \bibnamefont{Kavokin}},
  \bibinfo{journal}{Phys. Rev. B} \textbf{\bibinfo{volume}{64}},
  \bibinfo{pages}{075305} (\bibinfo{year}{2001}).

\bibitem[{\citenamefont{Kavokin}(2004)}]{kavokin2004:PRB}
\bibinfo{author}{\bibfnamefont{K.~V.} \bibnamefont{Kavokin}},
  \bibinfo{journal}{Phys. Rev. B} \textbf{\bibinfo{volume}{69}},
  \bibinfo{pages}{075302} (\bibinfo{year}{2004}).

\bibitem[{\citenamefont{Fabian et~al.}(2007)\citenamefont{Fabian,
  {Matos-Abiagus}, Ertler, Stano, and {\v{Z}uti\'{c}}}}]{fabian2007:APS}
\bibinfo{author}{\bibfnamefont{J.}~\bibnamefont{Fabian}},
  \bibinfo{author}{\bibfnamefont{A.}~\bibnamefont{{Matos-Abiagus}}},
  \bibinfo{author}{\bibfnamefont{C.}~\bibnamefont{Ertler}},
  \bibinfo{author}{\bibfnamefont{P.}~\bibnamefont{Stano}}, \bibnamefont{and}
  \bibinfo{author}{\bibfnamefont{I.}~\bibnamefont{{\v{Z}uti\'{c}}}},
  \bibinfo{journal}{Acta Phys. Slov.} \textbf{\bibinfo{volume}{57}},
  \bibinfo{pages}{565} (\bibinfo{year}{2007}).

\bibitem[{\citenamefont{Badescu et~al.}(2005)\citenamefont{Badescu,
  Lyanda-Geller, and Reinecke}}]{badescu2005:PRB}
\bibinfo{author}{\bibfnamefont{S.~C.} \bibnamefont{Badescu}},
  \bibinfo{author}{\bibfnamefont{Y.~B.} \bibnamefont{Lyanda-Geller}},
  \bibnamefont{and} \bibinfo{author}{\bibfnamefont{T.~L.}
  \bibnamefont{Reinecke}}, \bibinfo{journal}{Phys. Rev. B}
  \textbf{\bibinfo{volume}{72}}, \bibinfo{pages}{161304(R)}
  (\bibinfo{year}{2005}).

\bibitem[{\citenamefont{Glazov and Kulakovskii}(2009)}]{glazov2009:PRB}
\bibinfo{author}{\bibfnamefont{M.~M.} \bibnamefont{Glazov}} \bibnamefont{and}
  \bibinfo{author}{\bibfnamefont{V.~D.} \bibnamefont{Kulakovskii}},
  \bibinfo{journal}{Phys. Rev. B} \textbf{\bibinfo{volume}{79}},
  \bibinfo{pages}{195305} (\bibinfo{year}{2009}).

\bibitem[{\citenamefont{Stano and Fabian}(2006{\natexlab{a}})}]{stano2006:PRL}
\bibinfo{author}{\bibfnamefont{P.}~\bibnamefont{Stano}} \bibnamefont{and}
  \bibinfo{author}{\bibfnamefont{J.}~\bibnamefont{Fabian}},
  \bibinfo{journal}{Phys. Rev. Lett.} \textbf{\bibinfo{volume}{96}},
  \bibinfo{pages}{186602} (\bibinfo{year}{2006}{\natexlab{a}}).

\bibitem[{\citenamefont{Stano and Fabian}(2006{\natexlab{b}})}]{stano2006:PRB}
\bibinfo{author}{\bibfnamefont{P.}~\bibnamefont{Stano}} \bibnamefont{and}
  \bibinfo{author}{\bibfnamefont{J.}~\bibnamefont{Fabian}},
  \bibinfo{journal}{Phys. Rev. B} \textbf{\bibinfo{volume}{74}},
  \bibinfo{pages}{045320} (\bibinfo{year}{2006}{\natexlab{b}}).

\bibitem[{\citenamefont{Aleiner and Fa\v{l}ko}(2001)}]{aleiner2001:PRL}
\bibinfo{author}{\bibfnamefont{I.~L.} \bibnamefont{Aleiner}} \bibnamefont{and}
  \bibinfo{author}{\bibfnamefont{V.~I.} \bibnamefont{Fa\v{l}ko}},
  \bibinfo{journal}{Phys. Rev. Lett.} \textbf{\bibinfo{volume}{87}},
  \bibinfo{pages}{256801} (\bibinfo{year}{2001}).

\bibitem[{\citenamefont{Levitov and Rashba}(2003)}]{levitov2003:PRB}
\bibinfo{author}{\bibfnamefont{L.~S.} \bibnamefont{Levitov}} \bibnamefont{and}
  \bibinfo{author}{\bibfnamefont{E.~I.} \bibnamefont{Rashba}},
  \bibinfo{journal}{Phys. Rev. B} \textbf{\bibinfo{volume}{67}},
  \bibinfo{pages}{115324} (\bibinfo{year}{2003}).

\bibitem[{\citenamefont{Stano and Fabian}(2005)}]{stano2005:PRB}
\bibinfo{author}{\bibfnamefont{P.}~\bibnamefont{Stano}} \bibnamefont{and}
  \bibinfo{author}{\bibfnamefont{J.}~\bibnamefont{Fabian}},
  \bibinfo{journal}{Phys. Rev. B} \textbf{\bibinfo{volume}{72}},
  \bibinfo{pages}{155410} (\bibinfo{year}{2005}).

\bibitem[{\citenamefont{Wu and Lidar}(2003)}]{wu2003:PRL}
\bibinfo{author}{\bibfnamefont{L.-A.} \bibnamefont{Wu}} \bibnamefont{and}
  \bibinfo{author}{\bibfnamefont{D.~A.} \bibnamefont{Lidar}},
  \bibinfo{journal}{Phys. Rev. Lett.} \textbf{\bibinfo{volume}{91}},
  \bibinfo{pages}{097904} (\bibinfo{year}{2003}).

\bibitem[{\citenamefont{Kavokin}(2008)}]{kavokin2008:SST}
\bibinfo{author}{\bibfnamefont{K.~V.} \bibnamefont{Kavokin}},
  \bibinfo{journal}{Semicond. Sci. Technol.} \textbf{\bibinfo{volume}{23}},
  \bibinfo{pages}{114009} (\bibinfo{year}{2008}).

\bibitem[{\citenamefont{Golovach et~al.}(2008)\citenamefont{Golovach,
  Khaetskii, and Loss}}]{golovach2008:PRB}
\bibinfo{author}{\bibfnamefont{V.~N.} \bibnamefont{Golovach}},
  \bibinfo{author}{\bibfnamefont{A.}~\bibnamefont{Khaetskii}},
  \bibnamefont{and} \bibinfo{author}{\bibfnamefont{D.}~\bibnamefont{Loss}},
  \bibinfo{journal}{Phys. Rev. B} \textbf{\bibinfo{volume}{77}},
  \bibinfo{pages}{045328} (\bibinfo{year}{2008}).

\bibitem[{\citenamefont{Shen and Wu}(2007)}]{shen2007:PRB}
\bibinfo{author}{\bibfnamefont{K.}~\bibnamefont{Shen}} \bibnamefont{and}
  \bibinfo{author}{\bibfnamefont{M.~W.} \bibnamefont{Wu}},
  \bibinfo{journal}{Phys. Rev. B} \textbf{\bibinfo{volume}{76}},
  \bibinfo{pages}{235313} (\bibinfo{year}{2007}).

\bibitem[{\citenamefont{Sherman and Lockwood}(2005)}]{sherman2005:PRB}
\bibinfo{author}{\bibfnamefont{E.~Y.} \bibnamefont{Sherman}} \bibnamefont{and}
  \bibinfo{author}{\bibfnamefont{D.~J.} \bibnamefont{Lockwood}},
  \bibinfo{journal}{Phys. Rev. B} \textbf{\bibinfo{volume}{72}},
  \bibinfo{pages}{125340} (\bibinfo{year}{2005}).

\bibitem[{\citenamefont{Climente et~al.}(2007)\citenamefont{Climente, Bertoni,
  Goldoni, Rontani, and Molinari}}]{climente2007:PRB}
\bibinfo{author}{\bibfnamefont{J.~I.} \bibnamefont{Climente}},
  \bibinfo{author}{\bibfnamefont{A.}~\bibnamefont{Bertoni}},
  \bibinfo{author}{\bibfnamefont{G.}~\bibnamefont{Goldoni}},
  \bibinfo{author}{\bibfnamefont{M.}~\bibnamefont{Rontani}}, \bibnamefont{and}
  \bibinfo{author}{\bibfnamefont{E.}~\bibnamefont{Molinari}},
  \bibinfo{journal}{Phys. Rev. B} \textbf{\bibinfo{volume}{75}},
  \bibinfo{pages}{081303(R)} (\bibinfo{year}{2007}).

\bibitem[{\citenamefont{Olendski and Shahbazyan}(2007)}]{olendski2007:PRB}
\bibinfo{author}{\bibfnamefont{O.}~\bibnamefont{Olendski}} \bibnamefont{and}
  \bibinfo{author}{\bibfnamefont{T.~V.} \bibnamefont{Shahbazyan}},
  \bibinfo{journal}{Phys. Rev. B} \textbf{\bibinfo{volume}{75}},
  \bibinfo{pages}{041306(R)} (\bibinfo{year}{2007}).

\end{thebibliography}

\end{document}